\documentclass[cameraready]{Interspeech}

\title{VeRe-Flow: Guiding Flow Matching toward Clean Speech via Velocity Contrastive Regularization and Representation Alignment for Noise-Robust Bandwidth Expansion}

\author[affiliation={1}, orcid=0009-0006-2070-5253, equalcontribution]{Sujin}{Koo}
\author[affiliation={2}, equalcontribution]{Sangyoon}{Kim}
\author[affiliation={2}]{Ji Sub}{Um}
\author[affiliation={2}, correspondingauthor]{Hoirin}{Kim}

\address{
    $^1$ MAGO, South Korea \\
    $^2$ KAIST, South Korea
}

\email{sujin.koo@holamago.com, ndkim11@kaist.ac.kr, twiz0311@kaist.ac.kr, hoirkim@kaist.ac.kr}


\usepackage{comment} 
\usepackage{subcaption} 
\usepackage{booktabs} 
\usepackage{multirow} 
\usepackage{tikz} 
\usepackage{pgfplots} 
\pgfplotsset{compat=1.18} 
\usepackage{amsmath}
\usepackage{amssymb}
\usepackage{tabularx}
\usepackage{array}

\begin{document}

\maketitle

\begin{abstract}
Noise-robust bandwidth expansion aims to reconstruct high-fidelity wideband speech from noisy low-resolution inputs. While flow matching has shown strong performance in speech generation, accurately recovering clean speech from noisy inputs remains challenging due to the ambiguity of velocity estimation under noise. In this work, we propose \textbf{VeRe-Flow}, a clean-guided flow matching framework that introduces multi-level clean supervision to guide the generative process toward clean speech. At the velocity level, we introduce velocity contrastive regularization, which attracts the predicted velocity toward the clean trajectory while repelling it from noisy trajectories. At the representation level, we incorporate representation alignment that aligns intermediate features with clean self-supervised learning representations. The results demonstrate that the proposed method achieves the lowest LSD and highest DNSMOS OVRL among all baselines, and the highest MOS among generative baselines.

\end{abstract}

\footnotetext[1]{\url{https://vere-flow.github.io/VeRe-Flow-Demo/}}

\section{Introduction}

Noise-robust bandwidth expansion (NR-BWE) seeks to restore wideband speech from low-resolution inputs while suppressing environmental noise~\cite{UEE, MTL_MBE, EP-WUN, I_DTLN, SDNet, codec}. 
Low-resolution signals lack high-frequency components essential for naturalness and intelligibility, while background noise further degrades audio quality. Therefore, NR-BWE must simultaneously reconstruct missing spectral information and remain robust to noise.

Conventional BWE methods focus on recovering missing high-frequency bands~\cite{FLowHigh, lee2021nu, han2022nu, Liu_2022, liu2023audiosrversatileaudiosuperresolution, im2025flashsronestepversatileaudio}. However, they typically assume clean inputs and therefore degrade on noisy speech. Conversely, speech enhancement methods~\cite{lee2025flowse, richter2023speech} are effective at noise suppression but cannot reconstruct the missing spectral components. This highlights the need for models that can handle both bandwidth expansion and noise robustness.

Several studies have attempted to address this challenge~\cite{UEE, MTL_MBE, EP-WUN, I_DTLN, SDNet, codec}. For example, a recent codec-based model~\cite{codec} leverages quantized latent spaces for strong noise suppression. Despite these advances, existing approaches still struggle with the trade-off between accurate high-frequency reconstruction and effective noise suppression, leaving robust NR-BWE an open challenge.

Recently, flow matching-based generative models such as FLowHigh~\cite{FLowHigh} have shown strong performance for clean BWE, successfully achieving both high-frequency detail reconstruction and high perceptual audio quality. These properties make flow matching a promising framework for NR-BWE. However, the standard flow matching objective provides only one-sided supervision, encouraging the predicted velocity to follow the target direction. Under noisy conditions, this can lead to ambiguous velocity estimation and cause the generative trajectory to drift away from the clean speech manifold.

To address this, we propose \textbf{VeRe-Flow}, a clean-guided flow matching framework for NR-BWE that explicitly guides the generative process toward clean speech via \textbf{multi-level clean supervision} at both the velocity and representation levels. First, we introduce \textbf{velocity contrastive regularization} (VeCoR)~\cite{hong2025vecor} to provide bidirectional supervision in the velocity space. VeCoR attracts the predicted velocity toward the clean trajectory while repelling it from noisy trajectories, keeping the trajectory closer to the clean speech manifold. Second, a \textbf{representation alignment objective}~\cite{yurepa} aligns intermediate features with clean self-supervised learning (SSL) representations to promote noise-invariant semantic features. Finally, we enhance the backbone with convolutional residual modules and noise-robust SSL conditioning for additional semantic guidance. Audio samples are available\footnotemark[1].

Our main contributions are summarized as follows:

\begin{itemize}
    \item We introduce velocity contrastive regularization for NR-BWE, providing two-sided supervision that guides the predicted velocity toward the clean trajectory and away from noisy trajectories.
    \item We integrate a representation alignment objective that encourages intermediate features to capture clean semantic information via SSL representations.
    \item We integrate convolutional modules and noise-robust SSL conditioning into a unified flow-based NR-BWE framework.
    \item Experimental results show that the proposed method achieves the lowest LSD and highest DNSMOS OVRL among both generative and non-generative NR-BWE methods, and the best performance across LSD, all DNSMOS metrics, and MOS among generative models.
\end{itemize}

\section{Method}

\begin{figure*}[t]
    \centering
    \includegraphics[width=\textwidth]{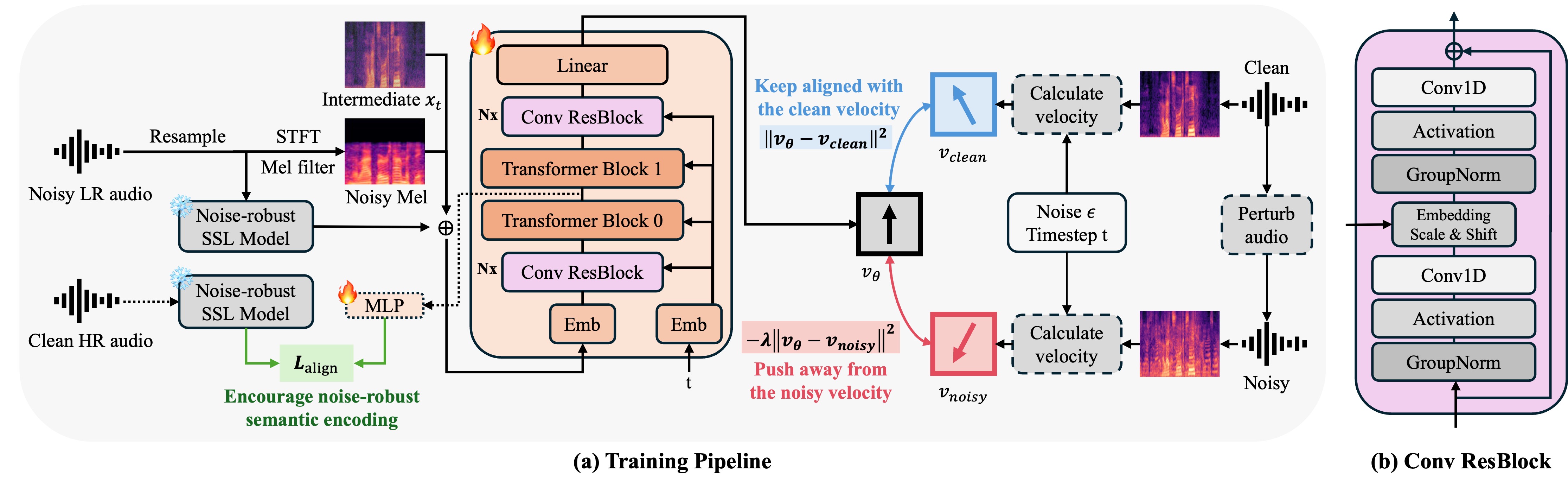}
    \caption{(a) Training process of the \textbf{VeRe-Flow}. The model predicts the velocity $v_\theta(x_t,t)$ conditioned on a noisy low-resolution mel-spectrogram and noise-robust self-supervised learning (SSL) features. \textbf{Velocity Contrastive Regularization} aligns $v_\theta$ with the clean trajectory and repels it from the noisy trajectories, guiding the predicted trajectory toward the clean speech manifold. Meanwhile, \textbf{Representation Alignment} aligns intermediate representations with clean SSL embeddings. (b) Architecture of the Conv ResBlock.}
    \label{fig:architecture}
\end{figure*}

\subsection{Preliminary: Conditional Flow Matching}
Flow matching~\cite{lipmanflow, tongimproving,pooladian2023multisample} is a generative modeling framework that learns a 
continuous velocity field transforming samples from a simple source 
distribution to a target distribution. Unlike diffusion models~\cite{ho2020denoising,songdenoising,songscore}, which 
generate samples through an iterative stochastic denoising process,  
flow matching directly parameterizes the velocity field of a 
deterministic probability flow.

Let $x \in \mathbb{R}^d$ denote a data point sampled from an unknown distribution $q(x)$. Flow matching defines a time-dependent probability density path $p_t(x)$ over $t \in [0,1]$, with $p_0$ being a known prior and $p_1$ approximating $q(x)$. The transformation is governed by the ODE:
\begin{equation}
\frac{d}{dt}\phi_t(x) = v_t(\phi_t(x)), \quad \phi_0(x) = x,
\label{eq:flow_ode}
\end{equation}
where $v_t : \mathbb{R}^d \times [0,1] \rightarrow \mathbb{R}^d$ is the target velocity field and $\phi_t(x)$ denotes the trajectory of sample $x$ over time $t$.

Directly optimizing this formulation is typically intractable, as it requires access to marginal densities and their corresponding velocity fields. Conditional Flow Matching(CFM) addresses this by conditioning on a target sample $x_1 \sim q(x)$, defining intermediate samples $x_t = \phi_t(x_0, x_1)$ along a path between source sample $x_0 \sim p_0$ and $x_1$, and introducing a conditional velocity field $u_t(x_t \mid x_1)$ that admits a closed-form solution under a linear interpolation path. The CFM objective is then:
\begin{equation}
\mathcal{L}_{\text{CFM}}(\theta) = \mathbb{E}_{t, q(x_1), p_t(x_t \mid x_1)} \left\Vert v_\theta(x_t, t) - u_t(x_t \mid x_1)\right\Vert^2,\label{eq:cfm_loss}
\end{equation}
where $v_\theta$ is a neural network parameterized by $\theta$ and $t \sim \mathcal{U}[0,1]$.

FLowHigh~\cite{FLowHigh} analyzed several conditional probability paths 
for clean bandwidth expansion. For the BWE task, they employed a data-dependent prior path using the LR mel-spectrogram as the source distribution, where $x_0 \sim \mathcal{N}(x_{LR}, I)$. In our NR-BWE task, we set $x_0 \sim \mathcal{N}(0, I)$ as the Gaussian prior and $x_1 = x_{HR}^{\mathrm{clean}}$ as the target high-resolution mel-spectrogram.  In our work, we  provide the noisy LR mel-spectrogram and SSL features as external conditions to the velocity network.

The interpolation path is given by:
\begin{equation}
x_t = (1 - (1 - \sigma_{\min})t)\, x_0 + t\, x_1,
\label{eq:linear_path}
\end{equation}
where $\sigma_{\min}$ controls the minimum noise level. The optimal conditional velocity field is obtained by differentiating Eq.~\eqref{eq:linear_path} with respect to $t$:
\begin{equation}
u_t^*(x_t \mid x_1) = x_1 - (1-\sigma_{\min})\,x_0,
\label{eq:optimal_velocity}
\end{equation}
which is constant with respect to $t$. Our model learns a conditioned velocity field $v_\theta(x_t, t \mid x_{LR}, f_{\text{SSL}}^{\text{noisy}})$ to approximate this optimal flow, where $f_{\text{SSL}}^{\text{noisy}}$ denotes SSL features extracted from the noisy low-resolution input.

\subsection{Model Architecture}

We incorporate 
DiC~\cite{tian2025dic}-style convolutional blocks adapted to 1D speech representations. As illustrated in 
Figure~\ref{fig:architecture}(b), each block follows a GroupNorm--Activation--Conv1D(kernel size 3) structure with mid-block scale-and-shift conditioning projected from 
the time embedding, followed by a residual connection.

As shown in Figure~\ref{fig:architecture}(a), we arrange these Conv ResBlocks and Transformer Blocks in a sandwich structure: a convolutional pre-stage, a central transformer stage, and a convolutional post-stage, where each convolutional stage consists of 4 stacked Conv ResBlocks. To provide noise-robust guidance, we incorporate frame-wise SSL features from frozen XEUS~\cite{chen2024towards}, which is pretrained with dereverberation and denoising objectives. The extracted representations $f_{\text{SSL}}^{\text{noisy}} \in \mathbb{R}^{T \times D}$ are obtained by encoding the noisy low-resolution input through XEUS after adjusting the temporal resolution, where $T$ and $D$ denote the number of frames and the feature dimension, respectively. These features are concatenated with $x_{LR}$ along the feature dimension and projected into the model input space. Except for these modifications, the overall architecture follows FLowHigh~\cite{FLowHigh}.

\subsection{Training Objectives}

\subsubsection{Representation Alignment Loss}
Since the model is conditioned on noisy input, its intermediate hidden states
may deviate from the clean speech manifold. Inspired by REPA~\cite{yurepa}, we introduce a representation alignment 
objective that encourages the model's intermediate hidden states to align with clean 
SSL representations during training.

Specifically, we extract the intermediate hidden state $h \in \mathbb{R}^{T \times d}$ 
from the output of the first transformer layer, where $d$ is the hidden dimension of the model, and project it into the SSL feature 
space via a 3-layer MLP projection head $\phi(\cdot)$ with SiLU activations:
\begin{equation}
    \mathcal{L}_{\text{align}} = \mathbb{E} \left[ -\sum_{t=1}^{T} 
    \text{sim}\left(f_{\text{SSL,t}}^{\text{clean}} ,\, \phi(h_t)\right) \right],
\end{equation}
where $f_{\text{SSL,t}}^{\text{clean}}$ denotes the frame-level XEUS SSL representation 
extracted from the clean high-resolution audio, and $\text{sim}(\cdot, \cdot)$ 
denotes cosine similarity. By minimizing this objective, the model is guided 
to produce internal representations consistent with clean speech characteristics, even when conditioned on degraded input.

\subsubsection{Velocity Contrastive Regularization Loss}
To explicitly guide the flow toward clean 
speech, we introduce velocity contrastive regularization (VeCoR), 
inspired by~\cite{hong2025vecor}.

Given $x_0 \sim \mathcal{N}(0, I)$, we define the clean and noisy 
target velocities as:
\begin{equation}
\begin{aligned}
u_t^{\mathrm{clean}} &= x_{HR}^{\mathrm{clean}} - (1-\sigma_{\min})\,x_0, \\
u_t^{\mathrm{noisy}} &= x_{HR}^{\mathrm{noisy}} - (1-\sigma_{\min})\,x_0,
\end{aligned}
\end{equation}
where $x_{HR}^{\mathrm{noisy}}$ is the mel-spectrogram of the semantically consistent
noise-perturbed high-resolution audio paired with $x_{HR}^{\mathrm{clean}}$.
The VeCoR objective attracts the predicted velocity toward the 
clean direction while repelling it from the noisy direction:
\begin{equation}
\mathcal{L}_{\mathrm{VeCoR}} = \mathbb{E}\left[
\lVert v_\theta - u_t^{\mathrm{clean}} \rVert^2
- \lambda_{\mathrm{VeCoR}} \lVert v_\theta - u_t^{\mathrm{noisy}} \rVert^2
\right]
\end{equation}
where $\lambda_{\mathrm{VeCoR}} > 0$ controls the strength of the repulsion. Note that the first term coincides with the CFM objective in Eq.~\eqref{eq:cfm_loss} evaluated at $x_1 = x_{HR}^{\mathrm{clean}}$.

\subsubsection{Overall Objective}
The final training objective combines $\mathcal{L}_{\mathrm{VeCoR}}$, which subsumes clean flow-matching supervision and contrastive repulsion from noisy velocities, with the representation alignment loss:
\begin{equation}
\mathcal{L}_{\mathrm{total}}
=
\mathcal{L}_{\mathrm{VeCoR}}
+
\lambda_{\mathrm{align}}
\mathcal{L}_{\mathrm{align}},
\end{equation}
where $\lambda_{\mathrm{align}}$ controls the strength of the representation alignment term.

\begin{table}[t]
\centering
\caption{
LSD, DNSMOS, and MOS results on the Valentini-Botinhao noisy test set downsampled to 8\,kHz.
Generative baselines marked with $\dagger$ are retrained as described in Section~\ref{subsec:baselines}.
}
\label{tab:dnsmos_lsd_results}

\fontsize{8}{9}\selectfont
\setlength{\tabcolsep}{3pt}
\renewcommand{\arraystretch}{1.1}

\begin{tabularx}{\linewidth}{l c c c c c c}
\toprule
\textbf{Method} & \textbf{NFE} & \textbf{LSD$\downarrow$} & \textbf{SIG$\uparrow$} & \textbf{BAK$\uparrow$} & \textbf{OVRL$\uparrow$} & \textbf{MOS$\uparrow$} \\
\midrule
GT & -- & 0.00 & 3.51 & 4.04 & 3.22 & 4.30$\pm$0.46 \\
GT recon & -- & 0.78 & 3.42 & 3.93 & 3.09 & --  \\
\midrule
\multicolumn{7}{l}{\textit{Non-generative models}} \\
\midrule
UEE~\cite{UEE}           & 1 & 2.72 & 2.27 & 2.39 & 2.17 & -- \\
MTL\_MBE~\cite{MTL_MBE} & 1 & 2.29 & 2.64 & 3.21 & 2.46 & -- \\
EP-WUN~\cite{EP-WUN} & 1 & 1.23 & \textbf{3.50} & 2.94 & 2.86 & -- \\
I-DTLN+~\cite{I_DTLN} & 1 & 1.54 & 2.63 & 2.87 & 2.18 & -- \\
SDNet~\cite{SDNet} & 1 & 1.16 & 3.29 & 3.32 & 2.92 & -- \\
Liu et al.~\cite{codec} & 1 & 1.54 & 3.28 & \textbf{4.08} & 3.04 & -- \\

\midrule
\multicolumn{7}{l}{\textit{Generative models}} \\
\midrule
NU-Wave2$^{\dagger}$~\cite{han2022nu} & 48 & 1.35 & 3.29 & 3.93 & 2.98 & 3.76$\pm$0.72 \\
FLowHigh$^{\dagger}$~\cite{FLowHigh} & 2 & \underline{1.12} & 3.40 & 3.91 & \underline{3.07} & 4.03$\pm$0.75 \\
\textbf{Proposed} & 2 & \textbf{1.10} & \underline{3.43} & \underline{3.97} & \textbf{3.12} & \textbf{4.14$\pm$0.65} \\
\bottomrule
\end{tabularx}
\end{table}

\footnotetext[2]{\url{https://huggingface.co/espnet/xeus}}
\footnotetext[3]{\url{https://github.com/hayeong0/Diff-HierVC}}
\footnotetext[4]{\url{https://github.com/microsoft/DNS-Challenge}}

\section{Experiments}
\label{sec:experiment}

\subsection{Dataset}

We conduct experiments on the Valentini-Botinhao dataset~\cite{valentini2017noisy}, a widely used benchmark for NR-BWE. It is a parallel clean–noisy corpus sampled at 48\,kHz, released in two subsets: a 28-speaker set and a 56-speaker set. To enhance speaker and data diversity, we merged both subsets, yielding a total of 84 speakers for training. For evaluation, we use the official test set, which contains two unseen speakers and 20 different noise conditions. To simulate bandwidth-limited inputs, we apply a Chebyshev Type-I low-pass filter followed by downsampling: during training, the filter order, ripple, and target sampling rate (1–15\,kHz) are randomly sampled, whereas evaluation signals are fixed at an order-8 filter with 0.05 dB ripple and downsampled to 8\,kHz, following the NR-BWE task setting. The reconstructed 16\,kHz outputs are compared against the corresponding 16\,kHz ground-truth references.
\subsection{Implementation Details}

All speech signals are converted into 80-dimensional mel-spectrograms with a 20\,ms hop size and a 1280-point window. Noise-robust SSL features are extracted using XEUS\footnotemark[2]~\cite{chen2024towards} every 20\,ms, aligned with the mel frame rate. The model is trained for 400k iterations with a batch size of 16 using the Adam optimizer~\cite{kingma2014adam} and a cosine annealing learning rate scheduler with a learning rate of $3\times10^{-4}$. During training, additive noise with a random SNR uniformly sampled from [5\,dB, 20\,dB] is applied with a probability of 0.5. The loss weights are set to $\lambda_{\text{align}}=0.25$ and $\lambda_{\text{VeCoR}}=0.05$. For waveform reconstruction, we employ the BigVGAN vocoder~\cite{lee2023bigvganuniversalneuralvocoder}.

\subsection{Baselines}
\label{subsec:baselines}

We compare the proposed model with both generative and non-generative noise-robust bandwidth expansion methods. As generative baselines, we consider FLowHigh~\cite{FLowHigh} and NU-Wave2~\cite{han2022nu} marked with $\dagger$ in the tables. Since these models were originally designed for clean bandwidth expansion, we retrain them under the same NR-BWE setting. Both FLowHigh and the proposed model generate mel-spectrograms with the same configuration and share a common vocoder, BigVGAN~\cite{lee2023bigvganuniversalneuralvocoder}, a publicly available pre-trained model\footnotemark[3] operating at 16\,kHz with 80 mel bins. In contrast, NU-Wave2 directly generates waveform signals and does not require a vocoder. For non-generative baselines, we report the results from the original papers, as in prior work, since their implementations are not publicly available.

\subsection{Evaluation Metrics}

For objective evaluation, we use Log-Spectral Distance (LSD) and Deep Noise Suppression Mean Opinion Score (DNSMOS\footnotemark[4])~\cite{reddy2022dnsmos}. LSD measures the spectral distance between reconstructed and reference signals, with lower values reflecting better performance. Meanwhile, DNSMOS is a neural non-intrusive speech quality estimator that objectively predicts the quality of speech (SIG), background noise (BAK), and overall quality (OVRL), where higher scores indicate better perceptual quality. While LSD is a standard metric for bandwidth expansion, DNSMOS is widely adopted to evaluate speech enhancement systems; both are reported as averages over the test set. For subjective evaluation, we conduct 5-point mean opinion score (MOS) tests on Amazon Mechanical Turk. We randomly select 40 test utterances and generate outputs from all compared generative models. Each sample is rated by 11  raters. The remaining evaluation protocol follows~\cite{babaev2024finally}.

\begin{table}[t]
\centering
\caption{
Effect of training strategy, ODE solver, and NFE on generative NR-BWE models. G and D denote Gaussian prior and Data-dependent prior, respectively, and Mid denotes the Midpoint ODE solver.
}
\label{tab:ablation_solver_training}
\fontsize{8}{10}\selectfont
\setlength{\tabcolsep}{2.9pt}
\renewcommand{\arraystretch}{1.1}
\begin{tabularx}{\linewidth}{l c c c c c c c}
\toprule
\textbf{Model} & \textbf{p$_0$} & \textbf{Solver} & \textbf{NFE} &
\textbf{LSD$\downarrow$} & \textbf{SIG$\uparrow$} &
\textbf{BAK$\uparrow$} & \textbf{OVRL$\uparrow$} \\
\midrule
\multirow{7}{*}{NU-Wave2$^{\dagger}$~\cite{han2022nu}}
& - & - & 2  & 5.64 & 2.04 & 1.31 & 1.31 \\
& - & - & 8  & 1.55 & 3.17 & 3.78 & 2.81 \\
& - & - & 16 & 1.66 & \underline{3.28} & \textbf{3.94} & \textbf{2.98} \\
& - & - & 32 & 1.41 & \textbf{3.29} & \underline{3.93} & \textbf{2.98} \\
& - & - & 48 & \underline{1.35} & \textbf{3.29} & \underline{3.93} & \textbf{2.98} \\
& - & - & 100 & \textbf{1.31} & \textbf{3.29} & 3.92 & \textbf{2.98} \\
\midrule
\multirow{6}{*}{FLowHigh$^{\dagger}$~\cite{FLowHigh}}
& D & Mid & 2 & 1.13 & 3.38 & 3.85 & 3.02 \\
& D & Euler & 2 & \textbf{1.11} & \underline{3.41} & \underline{3.89} & \underline{3.06} \\
& D & Euler & 6 & \textbf{1.11} & \textbf{3.42}  & 3.87 & \underline{3.06} \\
& G & Mid & 2 & 1.13 & 3.34 & 3.86 & 2.99 \\
& G & Euler & 2 & \underline{1.12} & 3.40 & \textbf{3.91} & \textbf{3.07} \\
& G & Euler & 6 & 1.13 & 3.40 & \underline{3.89} & 3.05 \\
\midrule
\multirow{6}{*}{\textbf{Proposed}}
& D & Mid & 2 & 1.18 & 3.41 & 3.94 & 3.09 \\
& D & Euler & 2 & \underline{1.11} & \underline{3.43} & \underline{3.96} & \underline{3.11} \\
& D & Euler & 6 & 1.16 & \textbf{3.44} & \underline{3.96} & \textbf{3.12} \\
& G & Mid & 2 & 1.17 & 3.38 & 3.93 & 3.06 \\
& G & Euler & 2 & \textbf{1.10} & \underline{3.43} & \textbf{3.97} & \textbf{3.12} \\
& G & Euler & 6 & 1.17 & 3.41 & \underline{3.96} & 3.10 \\
\bottomrule
\end{tabularx}
\end{table}

\begin{table}[ht]
\centering
\caption{Ablation studies on different SSL features at NFE=2.}
\label{tab:ssl_variant_ablation}
\resizebox{\columnwidth}{!}{%
\begin{tabular}{lcccc}
\toprule
\textbf{SSL} & \textbf{LSD ↓} & \textbf{SIG ↑} & \textbf{BAK ↑} & \textbf{OVRL ↑} \\
\midrule
XEUS~\cite{chen2024towards} (Proposed) & \textbf{1.10} & \textbf{3.43} & \textbf{3.97} & \textbf{3.12} \\
WavLM~\cite{chen2022wavlm}           & 1.15 & 3.42 & 3.94 & 3.10 \\
Wav2Vec 2.0~\cite{baevski2020wav2vec}    & 1.47 & 3.41 &  3.28 & 2.77  \\
\bottomrule
\end{tabular}
}
\end{table}

\section{Results}

\subsection{Baseline Comparisons}

As shown in Table~\ref{tab:ablation_solver_training}, we analyze the effect of the prior distribution, ODE solver, and number of function evaluations (NFE) on generative NR-BWE performance. Among the tested settings, the Gaussian prior with the Euler solver at 
NFE=2 achieves the best objective performance; we therefore adopt this configuration for all main comparisons in Table~\ref{tab:dnsmos_lsd_results}. Although the original FLowHigh paper reports results for clean BWE using a data-dependent prior with the midpoint solver at NFE=2, we re-evaluate FLowHigh under the same configuration to ensure a fair comparison with the proposed model. Since NU-Wave2~\cite{han2022nu} is a diffusion-based model that typically requires more function evaluations than flow matching models, we evaluate it at NFE=48, while its original paper reports results at NFE=8.

Table~\ref{tab:dnsmos_lsd_results} reports the objective and subjective results on the Valentini-Botinhao noisy test set downsampled to 8\,kHz. 
The proposed model achieves the best LSD and DNSMOS OVRL scores among all compared methods, indicating that it effectively improves both bandwidth expansion and noise suppression. In particular, among generative approaches, it outperforms the baselines across all objective metrics. The proposed model also obtains the highest MOS among the generative baselines, further demonstrating its perceptual advantage. Compared to FLowHigh, the proposed model consistently  yields higher DNSMOS scores across all tested configurations in Table~\ref{tab:ablation_solver_training}, indicating that the improvement is consistent across different training strategies and solver settings.

\begin{table}[t]
\centering
\caption{Ablation studies of the proposed model at NFE=2.}
\label{tab:ablation_adding}
\fontsize{8}{10}\selectfont
\setlength{\tabcolsep}{4pt}
\renewcommand{\arraystretch}{1.1}
\begin{tabularx}{\linewidth}{
c@{\hspace{3pt}}
>{\raggedright\arraybackslash}X
c
c
c
c
}
\toprule
\multicolumn{2}{l}{\textbf{Setting}} &
\textbf{LSD$\downarrow$} &
\textbf{SIG$\uparrow$} &
\textbf{BAK$\uparrow$} &
\textbf{OVRL$\uparrow$} \\
\midrule
(A):&FLowHigh$^{\dagger}$~\cite{FLowHigh} (Baseline)& 1.12 & 3.40 & 3.91 & 3.07 \\
\midrule
(B):& (A) + Conv ResBlock& 1.11 & 3.42 & 3.91 & 3.08 \\
(C):& (B) + XEUS& \textbf{1.08} & 3.41 & 3.94 & 3.09 \\
(D):& (C) + REPA& 1.09 & \textbf{3.43} & 3.94 & 3.11 \\
(E):& (D) + VeCoR (Proposed)& 1.10 & \textbf{3.43} & \textbf{3.97} & \textbf{3.12} \\
\midrule
(F):& (E) - Conv ResBlock& 1.09 & 3.41 & 3.96 & 3.10 \\
\bottomrule
\end{tabularx}
\end{table}

\subsection{Ablation on Choice of SSL Features}
Table~\ref{tab:ssl_variant_ablation} shows the effect of replacing XEUS~\cite{chen2024towards} with alternative self-supervised features. XEUS~\cite{chen2024towards} performs best in terms of both LSD and DNSMOS, outperforming WavLM~\cite{chen2022wavlm} and Wav2Vec 2.0~\cite{baevski2020wav2vec} in reducing spectral distortion and improving perceptual audio quality. This indicates that XEUS is the most suitable choice for NR-BWE.

\subsection{Ablation on Model Components}

Table~\ref{tab:ablation_adding} shows the contribution of each component by progressively adding modules to the base model. Conv ResBlock improves SIG and OVRL, indicating enhanced overall audio quality. XEUS yields the largest reduction in LSD with improved BAK and OVRL, suggesting that its noise-robust SSL representation improves spectral reconstruction for bandwidth expansion. REPA improves SIG and OVRL by aligning intermediate representations with clean speech features, while VeCoR contributes to improved BAK and OVRL, indicating better robustness to background noise.
Overall, XEUS primarily improves LSD for bandwidth expansion, while REPA and VeCoR improve DNSMOS scores for speech enhancement. Together, these components enable effective NR-BWE. Additionally, removing the Conv ResBlock from the full model still outperforms the baseline across all metrics, confirming that the remaining components are effective for noise-robust bandwidth expansion even without the modified backbone, relative to the original FLowHigh~\cite{FLowHigh} architecture.

\section{Conclusion}
We propose VeRe-Flow, a flow matching framework for NR-BWE. It introduces a clean-targeted supervision strategy that regularizes the generative process at both the velocity and representation levels. Velocity contrastive regularization encourages the predicted velocity field to follow the clean speech manifold while repelling noisy directions, whereas representation alignment guides intermediate representations to better capture clean speech characteristics. To the best of our knowledge, this work is the first to apply velocity contrastive regularization to speech generation. Experiments on the Valentini-Botinhao noisy benchmark demonstrate that the proposed model outperforms existing generative baselines across all metrics, achieving lower LSD and higher DNSMOS and MOS scores.

\ifcameraready
     \section{Acknowledgment}
     This work was supported by Institute of Information \& communications Technology Planning \& Evaluation (IITP) grant funded by the Korea government(MSIT) (No.RS-2025-02215393).
\else
     
\fi

\section{Generative AI Use Disclosure}
Generative AI tools were used to assist in editing and improving the clarity of the manuscript. The technical content, experimental design, analysis, and conclusions are entirely the work of the authors.

\bibliographystyle{IEEEtran}
\bibliography{refs}

@inproceedings{lee2021nu,
  title={NU-Wave: A Diffusion Probabilistic Model for Neural Audio Upsampling},
  author={Lee, Junhyeok and Han, Seungu},
  booktitle={Interspeech},
  year={2021}
}

@inproceedings{han2022nu,
  title={NU-Wave 2: A General Neural Audio Upsampling Model for Various Sampling Rates},
  author={Han, Seungu and Lee, Junhyeok},
booktitle={Interspeech},
  year={2022}
}

@inproceedings{Liu_2022,
   title={Neural Vocoder is All You Need for Speech Super-resolution},
   booktitle={Interspeech}, 
year={2022},
   author={Liu, Haohe and Choi, Woosung and Liu, Xubo and Kong, Qiuqiang and Tian, Qiao and Wang, DeLiang},
 }

@inproceedings{liu2023audiosrversatileaudiosuperresolution,
  title={AudioSR: Versatile Audio Super-resolution at Scale}, 
  author={Haohe Liu and Ke Chen and Qiao Tian and Wenwu Wang and Mark D. Plumbley},
  year={2024},
booktitle={International Conference on Acoustics, Speech, and Signal Processing},
}

@inproceedings{im2025flashsronestepversatileaudio,
  title={FlashSR: One-step Versatile Audio Super-resolution via Diffusion Distillation}, 
  author={Jaekwon Im and Juhan Nam},
booktitle={Interspeech},
  year={2025},
}

@inproceedings{flowhigh,
  title={FLowHigh: Towards Efficient and High-Quality Audio Super-Resolution with Single-Step Flow Matching}, 
  author={Jun-Hak Yun and Seung-Bin Kim and Seong-Whan Lee},
  year={2025},
booktitle={International Conference on Acoustics, Speech, and Signal Processing},
}

@inproceedings{chen2024towards,
  title={Towards Robust Speech Representation Learning for Thousands of Languages},
  author={Chen, William and Zhang, Wangyou and Peng, Yifan and Li, Xinjian and Tian, Jinchuan and Shi, Jiatong and Chang, Xuankai and Maiti, Soumi and Livescu, Karen and Watanabe, Shinji},
  year={2024},
  booktitle={Empirical Methods in Natural Language Processing},
}

@article{ho2020denoising,
  title={Denoising diffusion probabilistic models},
  author={Ho, Jonathan and Jain, Ajay and Abbeel, Pieter},
  journal={Advances in Neural Information Processing Systems},
  year={2020}
}

@inproceedings{songdenoising,
  title={Denoising Diffusion Implicit Models},
  author={Song, Jiaming and Meng, Chenlin and Ermon, Stefano},
  booktitle={International Conference on Learning Representations},
  year={2021},
}

@inproceedings{lipmanflow,
  title={Flow Matching for Generative Modeling},
  author={Lipman, Yaron and Chen, Ricky TQ and Ben-Hamu, Heli and Nickel, Maximilian and Le, Matthew},
  booktitle={International Conference on Learning Representations},
  year={2023}
}

@inproceedings{lee2023bigvganuniversalneuralvocoder,
  title={BigVGAN: A Universal Neural Vocoder with Large-Scale Training}, 
  author={Sang-gil Lee and Wei Ping and Boris Ginsburg and Bryan Catanzaro and Sungroh Yoon},
  booktitle={International Conference on Learning Representations},
  year={2023}
}

@article{valentini2017noisy,
  title={Noisy reverberant speech database for training speech enhancement algorithms and tts models},
  author={Valentini-Botinhao, Cassia},
  year={2017},
  journal={University of Edinburgh, School of Informatics, Centre for Speech Technology Research (CSTR)}
}

@inproceedings{reddy2022dnsmos,
  title={DNSMOS P. 835: A non-intrusive perceptual objective speech quality metric to evaluate noise suppressors},
  author={Reddy, Chandan KA and Gopal, Vishak and Cutler, Ross},
  booktitle={International Conference on Acoustics, Speech, and Signal Processing},
year={2022}
}

@article{baevski2020wav2vec,
  title={wav2vec 2.0: A framework for self-supervised learning of speech representations},
  author={Baevski, Alexei and Zhou, Yuhao and Mohamed, Abdelrahman and Auli, Michael},
  journal={Advances in Neural Information Processing Systems},
  year={2020}
}

@article{chen2022wavlm,
  title={Wavlm: Large-scale self-supervised pre-training for full stack speech processing},
  author={Chen, Sanyuan and Wang, Chengyi and Chen, Zhengyang and Wu, Yu and Liu, Shujie and Chen, Zhuo and Li, Jinyu and Kanda, Naoyuki and Yoshioka, Takuya and Xiao, Xiong and others},
  journal={IEEE Journal of Selected Topics in Signal Processing},
  year={2022},
}

@INPROCEEDINGS{SDNet,
  author={Yang, Junkang and Liu, Hongqing and Gan, Lu and Zhou, Yi and Li, Xing and Jia, Jie and Yao, Jinzhuo},
  booktitle={Asia Pacific Signal and Information Processing Association Annual Summit and Conference}, 
year={2024},
  title={SDNet: Noise-Robust Bandwidth Extension under Flexible Sampling Rates}, 
}

@inproceedings{codec,
title={A Neural Codec Approach for Noise-Robust Bandwidth Expansion},
  author={Liu, Xi and Yang, Mu and Chen, Szu-Jui and Hansen, John HL},
  booktitle={Interspeech},
  year={2025}
}

@INPROCEEDINGS{UEE,
  author={Liu, Bin and Tao, Jianhua and Zheng, Yibin},
  booktitle={International Symposium
on Chinese Spoken Language Processing},
year = {2018}, 
  title={A Novel Unified Framework for Speech Enhancement and Bandwidth Extension Based on Jointly Trained Neural Networks}, 
  }

@inproceedings{MTL_MBE,
author = {Hou, Nana and Xu, Chenglin and Zhou, Joey and Chng, Eng and Li, Haizhou},
year = {2020},
title = {Multi-Task Learning for End-to-End Noise-Robust Bandwidth Extension},
booktitle = {Interspeech},
}

@inproceedings{EP-WUN,
author = {Lin, Yin-Tse and Su, Bo and Lin, Chi-Han and Kuo, Shih-Chan and Jang, Jyh-Shing and Lee, Chi-Chun},
booktitle = {Interspeech},
year= {2023},
title = {Noise-Robust Bandwidth Expansion for 8K Speech Recordings},
}

@INPROCEEDINGS{I_DTLN,
  author={Chen, Che-Wen and Wang, Wei-Chun and Ou, Yang-Yen and Wang, Jhing-Fa},
  booktitle={International
Conference on Orange Technology}, 
  title={Deep Learning Audio Super Resolution and Noise Cancellation System for Low Sampling Rate Noise Environment},
year={2022}
}

@inproceedings{songscore,
  title={Score-Based Generative Modeling through Stochastic Differential Equations},
  author={Song, Yang and Sohl-Dickstein, Jascha and Kingma, Diederik P and Kumar, Abhishek and Ermon, Stefano and Poole, Ben},
  booktitle={International Conference on Learning Representations},
year={2021}
}

@inproceedings{tian2025dic,
  title={Dic: Rethinking conv3x3 designs in diffusion models},
  author={Tian, Yuchuan and Han, Jing and Wang, Chengcheng and Liang, Yuchen and Xu, Chao and Chen, Hanting},
  booktitle={Proceedings of the Computer Vision and Pattern Recognition Conference},
  year={2025}
}

@InProceedings{hong2025vecor,
    author    = {Hong, Zong-Wei and Li, Jing-Lun and Li, Lin-Ze and Zhang, Shen and Tang, Yao},
    title     = {VeCoR-Velocity Contrastive Regularization for Flow Matching},
    booktitle = {Proceedings of the IEEE/CVF Conference on Computer Vision and Pattern Recognition (CVPR) Findings},
    year      = {2026},
}

@article{babaev2024finally,
  title={FINALLY: fast and universal speech enhancement with studio-like quality},
  author={Babaev, Nicholas and Tamogashev, Kirill and Saginbaev, Azat and Shchekotov, Ivan and Bae, Hanbin and Sung, Hosang and Lee, WonJun and Cho, Hoon-Young and Andreev, Pavel},
  journal={Advances in Neural Information Processing Systems},
  year={2024}
}

@inproceedings{yurepa,
  title={Representation Alignment for Generation: Training Diffusion Transformers Is Easier Than You Think},
  author={Yu, Sihyun and Kwak, Sangkyung and Jang, Huiwon and Jeong, Jongheon and Huang, Jonathan and Shin, Jinwoo and Xie, Saining},
  booktitle={International Conference on Learning Representations},
  year={2025}
}

@article{tongimproving,
  title={Improving and generalizing flow-based generative models with minibatch optimal transport},
  author={Tong, Alexander and FATRAS, Kilian and Malkin, Nikolay and Huguet, Guillaume and Zhang, Yanlei and Rector-Brooks, Jarrid and Wolf, Guy and Bengio, Yoshua},
  journal={Transactions on Machine Learning Research},
  year={2024}
}

@inproceedings{pooladian2023multisample,
  title={Multisample Flow Matching: Straightening Flows with Minibatch Couplings},
  author={Pooladian, Aram-Alexandre and Ben-Hamu, Heli and Domingo-Enrich, Carles and Amos, Brandon and Lipman, Yaron and Chen, Ricky TQ},
  booktitle={International Conference on Machine Learning},
  year={2023},
}

@inproceedings{lee2025flowse,
  title={Flowse: Flow matching-based speech enhancement},
  author={Lee, Seonggyu and Cheong, Sein and Han, Sangwook and Shin, Jong Won},
  booktitle={International Conference on Acoustics, Speech, and Signal Processing},
  year={2025},
}

@article{richter2023speech,
  title={Speech enhancement and dereverberation with diffusion-based generative models},
  author={Richter, Julius and Welker, Simon and Lemercier, Jean-Marie and Lay, Bunlong and Gerkmann, Timo},
  journal={IEEE/ACM Transactions on Audio, Speech, and Language Processing},
  year={2023},
}

@article{kingma2014adam,
  title={Adam: A method for stochastic optimization},
  author={Kingma, Diederik P and Ba, Jimmy},
  journal={arXiv preprint arXiv:1412.6980},
  year={2014}
}

\end{document}